\documentclass[aps,pra,twocolumn,superscriptaddress]{revtex4}
\usepackage{amsfonts}
\usepackage{amssymb}
\usepackage{amsmath}
\usepackage{epsfig}
\usepackage{color}
\usepackage{graphics, graphicx}
\usepackage{bbold}
\usepackage{psfrag}
\usepackage{mathcomp}
\usepackage{subfigure}
\usepackage{verbatim}
\usepackage{color}
\usepackage[colorlinks,citecolor=blue]{hyperref}

\usepackage{soul}
\usepackage{ulem}
\usepackage{cancel}

\begin{document}

\title{Exploring interacting topological insulator of extended Su-Schrieffer-Heeger model}
\author{Xiaofan Zhou}
\affiliation{State Key Laboratory of Quantum Optics and Quantum Optics Devices, Institute
of Laser spectroscopy, Shanxi University, Taiyuan 030006, China}
\affiliation{Collaborative Innovation Center of Extreme Optics, Shanxi University,
Taiyuan, Shanxi 030006, China}
\author{Jian-Song Pan}
\email{panjsong@scu.edu.cn}
\affiliation{College of Physics, Sichuan University, Chengdu 610065, China}
\affiliation{Key Laboratory of High Energy Density Physics and Technology of Ministry of Education, Sichuan University, Chengdu 610065, China}
\author{Suotang Jia}
\affiliation{State Key Laboratory of Quantum Optics and Quantum Optics Devices, Institute
of Laser spectroscopy, Shanxi University, Taiyuan 030006, China}
\affiliation{Collaborative Innovation Center of Extreme Optics, Shanxi University,
Taiyuan, Shanxi 030006, China}

\begin{abstract}
Exploring topological phases in interacting systems is a challenging task.
We investigate many-body topological physics of interacting fermions in an extended
Su-Schrieffer-Heeger (SSH) model, which extends the two sublattices of SSH model into four sublattices and thus is dubbed SSH4 model, based on the density-matrix
renormalization-group numerical method.
The interaction-driven phase transition from topological insulator to charge density wave (CDW) phase can be identified by analyzing the variations of entanglement spectrum, entanglement entropies, energy gaps, CDW order parameter, and fidelity.
We map the global phase diagram of the many-body ground state, which contains nontrivial topological insulator, trivial insulator and CDW phases, respectively. In contrast to interacting SSH model, in which the phase transitions to the CDW phase are argued to be first-order phase transitions, the phase transitions between the CDW phase and topologically trivial/nontrivial phases are shown to be continuous phase transitions.
Finally, we {also} show the phase diagram of interacting spinful SSH4 model, where the attractive (repulsive) on-site spin interaction amplifies (suppresses) the CDW phase.
 The models analyzed here can be implemented with ultracold atoms on optical superlattices.
\end{abstract}

\maketitle

\section{Introduction}
\label{Introduction}

Understanding the topological properties of band insulator and interacting topological insulator
is one of the most fundamental and challenging tasks in the studies of condensed matter materials and ultracold atomic gases~\cite{Hasan2010,Qi2011,Chiu2016,Bansil2016,Goldman2016,Armitage2018,zhang2018,Cooper2019}.
As a highly controllable and disorder-free system, ultracold atoms in optical lattices provide
a powerful platform for quantum simulation of topological states of matter~\cite{Goldman2016,zhang2018,Cooper2019}.
One of the most basic and easiest models in describing band topology is the celebrated Su-Schrieffer-Heeger (SSH) model~\cite{SSH1979},
which has been experimentally implemented with ultracold atoms in one-dimensional (1D) dimerized optical
superlattices~\cite{Atala2013,Wang2013,Lohse2016,Nakajima2016,Leder2016}.

The SSH model describes noninteracting quantum particles hopping in a 1D lattice
with alternating hopping coefficients.
Varying the hopping ratio, the topological trivial phase or nontrivial phase appears,
depending on the hopping term on the end of SSH model is strong or weak~\cite{shen2012}.
For a noninteracting topological insulator, edge
degeneracy comes directly from the zero-energy edge mode,
which is protected by its topological invariants of the bulk crystal through the
bulk-edge correspondence.
After considering the interaction, the SSH model exhibits a rich phase diagram~\cite{Yoshida2014,Sirker2014,Wang2015,Liberto2016,Ye2016,Meichanetzidis2016,Marques2017,Kuno2019} and interesting topological bound states~\cite{marques2018topological}, where the single-particle picture is not applicable.

On the other hand, stimulated by experimental progresses, many variations and extensions of the SSH model have been proposed and explored, such as driven SSH model~\cite{GomezLeon2013,DalLago2015}, SSH model with long-range hopping~\cite{Li2014,An2018,PerezGonzalez2019,Ahmadi2020,kumar2021}, two-leg SSH model~\cite{zhangtwolegSSH}, Creutz ladder model~\cite{Sun2017,Junemann2017,Kang2018,He2021} and extended SSH model~\cite{SSH42018}.
One typical extended example is to change the site period of the unit cell from two to
four, thus one can transform the standard SSH model into the
considerably richer SSH4 model with four hopping coefficients~\cite{SSH42018}.
The wider parameter space of the SSH4 model is useful for studying
topological properties of system with higher dimensions including synthetic dimension~\cite{Celi2014,Price2015,Mancini2015,Stuhl2015,Lustig2019}.
The SSH4 model has the chiral symmetry and belongs
to the same topological class of the SSH model, and the winding number
can characterize its band topology~\cite{SSH42018}.
With open boundary condition, there exist topological edge states at the boundary
of the system~\cite{marques2020analytical}.
For a SSH4 model with infinite sites, the topological trivial and nontrivial phases are determined by
the tunneling ratio.
So far, the single-particle topological characterizations of SSH4 have been investigated clearly~\cite{SSH42018,SSH42019,SSH42020}.
However, to the best of our knowledge, a detailed study of interacting SSH4 model is still lacked.

In this paper, we investigate interacting topological properties of spinless and spin-1/2 SSH4 models in 1D optical superlattices, based on density-matrix renormalization-group (DMRG) numerical method~\cite{dmrg1,dmrg2}.
For interacting SSH4 model, the topological invariant and classification of interacting TI become $\mathbb{Z}_{4}$, which are different from the single-particle TI classified with $\mathbb{Z}$ group.
The nearest-neighbor interaction can drive the topological insulator (TI) and the topologically trivial insulator phases to the charge density wave (CDW) phase, which is characterized with the entanglement spectrum, entanglement entropies, energy gaps, and CDW order parameter. We numerically work out the many-body phase diagrams, and show the typical features of the appeared quantum phases. Although the phase diagram is similar to that of interacting SSH model, we find the phase transitions to the CDW phase are continuous phase transitions, unlike those in the SSH model, which are argued to be first-order phase transitions based on variational study~\cite{Yahyavi2018}. The central charges at the phase boundaries between the CDW and TI/trivial insulator (TI and trivial insulator) phase are shown to be 2 (1). Further, we analyze the ground states of interacting spinful SSH4 model.
It shows that the repulsive on-site interaction of spin-1/2 SSH4 model can enhance the TI phase and suppress the
CDW phase, but the attractive on-site interaction plays the opposite role. Our results may stimulate a new avenue for simulating interacting fermionic topological phases using cold atom in optical lattices.

\section{Model and Hamiltonian}
\label{Model and Hamiltonian}

\begin{figure}[t]
\centering
\includegraphics[width = 3.5in]{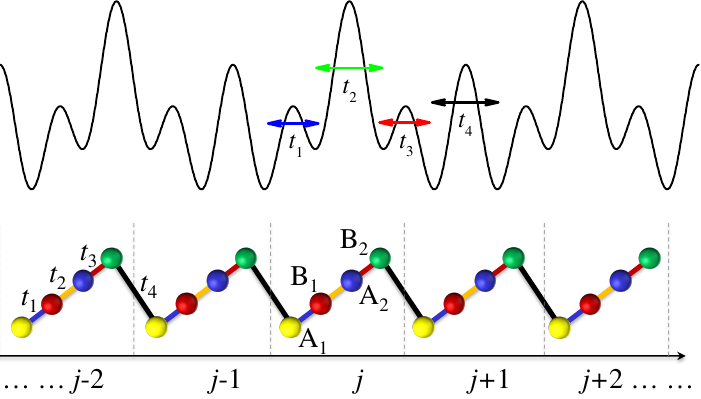} \hskip 0.0cm
\caption{The sketch of the SSH4 model. Three superposed optical lattices
with lattice constant $a/2$, $a$ and $2a$ effectively realize an SSH4 model.
This model exhibits four sites per unit cell with four tunnelings, which can be tuned independently by varying the three lattice strengths and
relative phase between the lattices.}
\label{fig:sketch}
\end{figure}

In cold atom experiment, three superposed optical lattices with lattice constant $a/2$, $a$ and $2a$
effectively realize an SSH4 model, as shown in
Fig.~\ref{fig:sketch}. The three optical lattices may be obtained from a single
laser working at $\lambda_{\mathtt{laser}} = 1064 $nm.
The $a/2$ optical lattice can
be obtained by retroreflecting the frequency-doubled laser, with $a =
\lambda_{\mathtt{laser}}/2$.
The $a$ lattice may be obtained by
retroreflecting the laser $\lambda_{\mathtt{laser}}$.
The lattice at $2a$
may be obtained by crossing two $\lambda_{\mathtt{laser}}$ beams with a small
angle~\cite{SSH42018}.
This superlattice exhibits four sites per unit cell (see Fig.~\ref{fig:sketch}), and hence is called as SSH4 model.
The tight-binding interacting SSH4 Hamiltonian can be written as
\begin{eqnarray}
H &=&\sum_{j=1}^{L/4}[t_{1}\hat{c}_{4j-3}^{\dag }\hat{c}_{4j-2}+t_{2}\hat{c}%
_{4j-2}^{\dag }\hat{c}_{4j-1}+t_{3}\hat{c}_{4j-1}^{\dag }\hat{c}_{4j}  \notag
\\
&&+t_{4}\hat{c}_{4j}^{\dag }\hat{c}_{4j+1}+\text{H.c.}]+V\sum_{j}^{L}\hat{n}%
_{j}\hat{n}_{j+1},
\label{H1}
\end{eqnarray}
where $\hat{c}_{j}$ ($\hat{c}_{j}^{\dag }$) are fermionic annihilation
(creation) operators of the $j$th site, $\hat{n}_{j}=\hat{c}_{j}^{\dag }%
\hat{c}_{j}$,
$t_n$ are the tunneling rates, and $V$ measures the nearest-neighbor density-density interaction.
In this configuration, $t_{1}=t_{3}$, but $t_{2}$ and $t_{4}$
can be tuned independently by varying the three lattice strengths and
relative phase between the three lattices.

For single-particle items of Hamiltonian (\ref{H1}), the
time-reversal, particle-hole, and chiral symmetries exist, then the
topological insulator belongs to the symmetry class BDI of the
Altland-Zirnbauer classification and is characterized by a $\mathbb{Z}$
invariant~\cite{Altland1997,Schnyder2008,Ludwig2016,Chiu2016}.
When $t_{1}=t_{3}$ and $t_{2}=t_{4}$, the SSH4 model reduces to the common SSH model.
The SSH4 model has four bands with more mid-gap states located inside the three gaps.
However, only the zero-energy state is protected by the chiral symmetry and associated with the band-topology.
Thus, the winding number is defined to identify the property of the two negative (or positive) energy bands, which all contribute to its value.
The winding number $w=1$ when $|t_{1}t_{3}|<|t_{2}t_{4}|$, and $w=0$ when $|t_{1}t_{3}|>|t_{2}t_{4}|$.
In the presence of weak interaction $V$, the topological invariant become $\mathbb{Z}%
_{4}$~\cite{Morimoto2015}.
In the following, we focus on the two negative topology bands with zero-energy states, which corresponds to the half-filling occupation, i.e., $N/L=0.5$, with the atom number $N$ and lattice length $L$.

The Hamiltonian (\ref{H1}) shows the spinless SSH4 model. In cold atoms
experiment, the hyperfine states of the atoms usually can be
treated as components of spin. After considering the two hyperfine
states, we can get the spin-1/2 SSH4 Hamiltonian, which can be written as
\begin{eqnarray}
H &=&\sum_{j=1,\sigma }^{L/4}[t_{1}\hat{c}_{4j-3,\sigma }^{\dag }\hat{c}%
_{4j-2,\sigma }+t_{2}\hat{c}_{4j-2,\sigma }^{\dag }\hat{c}_{4j-1,\sigma }
\notag \\
&&+t_{3}\hat{c}_{4j-1,\sigma }^{\dag }\hat{c}_{4j,\sigma }+t_{4}\hat{c}%
_{4j,\sigma }^{\dag }\hat{c}_{4j+1,\sigma }+\text{H.c.}]  \notag \\
&&+V\sum_{j}\hat{n}_{j}\hat{n}_{j+1}+U\sum_{j}\hat{n}_{j,\uparrow }\hat{n}%
_{j,\downarrow },
\label{H2}
\end{eqnarray}
where $\sigma$ presents the spin-up and spin-down, $\uparrow,\downarrow$ for spin-1/2 fermion,
$U$ is a on-site interaction strength between opposite spin due to $s$-wave scattering with $\hat{n}_{j} = \sum_{\sigma} \hat{c}_{j,\sigma}^{\dag }\hat{c}%
_{j,\sigma}$ the number operator. With the spin degree of freedom, the model exhibits eight energy bands.
The topological insulator is presented at the filling $N/L=1$.

In order to quantitatively reveal the SSH4 models, we will perform DMRG numerical method with lattice length up to $L=320$ for spinless SSH4 model and
$L=128$ for spin-1/2 SSH4 model, for which we retain 400 truncated states per DMRG
block and perform 30 sweeps with acceptable truncation errors~\cite{DMRG_details}.

\begin{figure*}[t]
\centering
\includegraphics[width = 7.0in]{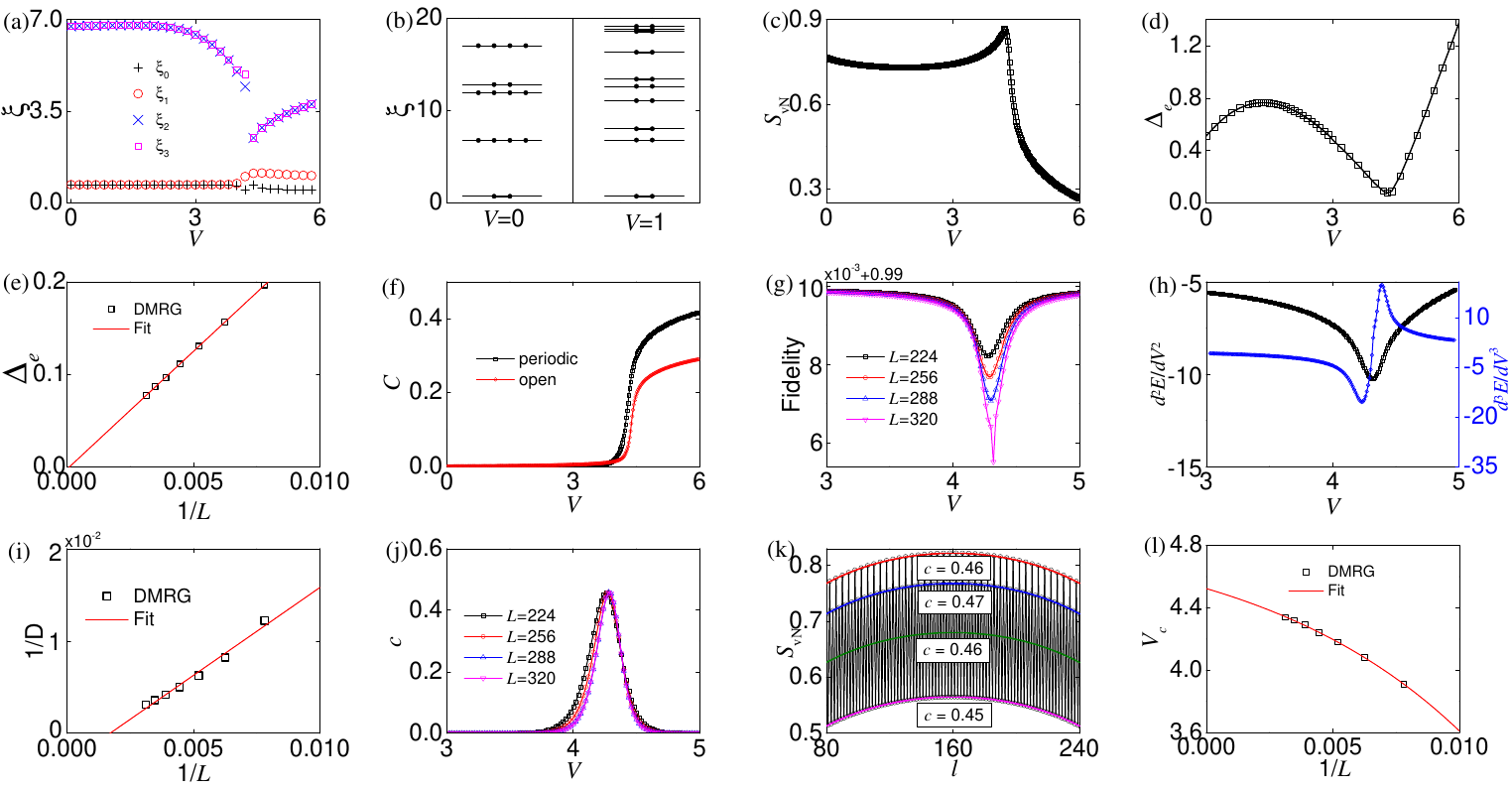} \hskip 0.0cm
\caption{(a) The lowest four levels in the entanglement spectrum $\protect%
\xi_{i}$ ($i=0,1,2,3$) as a function of the interaction strength $V$.
(b) The 16 lower levels in the entanglement spectrum for two values of $V=0,1$.
(c) The von Neumann entropy $S_{\mathrm{vN}}$, and
(d) the excited energy gap $\Delta_{e}$ versus $V$.
(e) The finite-size scaling of the $\Delta_{e}$ at the critical point, and the red solid line is a linear fit with $\Delta_{e} \sim 0$ in the large-$L$ limit.
(f) The CDW order parameter $C$,
(g) fidelity $\langle \psi(V) | \psi(V+\delta V) \rangle$ ($\delta V=0.02$) of finite lattice lengths, and
(h) the derivatives of ground-state energy $d^{n}E/dV^{n}$ ($n=2,3$) versus $V$.
(i) The finite-size scaling of the inverse of peak value of the fourth-order derivative of ground-state energy (not show here)], and
(j) The fitting central charge $c$ (note that only at the critical point $c$ can be defined, although we always can fit the formula and obtain a value) as functions of the interaction strength $V$.
(k) The scaling of the von Neumann entropy $S_{\mathrm{vN}}(l)$ as a function of subchain $l$ at the critical point.
The green line is $S_{\mathrm{vN}}(l)=\frac{c}{3}\ln[\sin(\protect\pi l/L)]+\text{const}$ with $c=0.46$ (extracted from the fitting of the mean values of $S_{\mathrm{vN}}$).
(l) The finite-size scaling of the critical point for phase transition between TI and CDW with $t_2=1.6$. The critical point $V_c=4.53$ in the thermodynamic limits.
In all subfigure, we have $t_2=1.6$ and $L=320$ except the finite-size scalings. All subfigure are under periodic boundary condition.}
\label{fig:phase transition}
\end{figure*}


\section{Order parameters}
\label{order parameters}

The strong-correlated topological properties can be well described by the degeneracy
in entanglement spectrum of ground-state, entanglement entropy, and excited energy gap.
The system is topological nontrivial if the
entanglement spectrum is degenerate since the entanglement spectrum
is associated with the energy spectrum of edge excitations \cite%
{Zhao2015,Yoshida2014,Turner2011,Pollmann2010,
Fidkowski2010,Flammia2009,Li2008}.
The entanglement spectrum is defined as a logarithmic rescaling of the Schmidt
values~\cite{Li2008}%
\begin{equation}
\xi _{i}=-\ln (\rho _{i}),
\end{equation}%
with $\rho_{i}$ being the eigenvalue of the reduced density matrix $\hat{%
\rho}_{l}=\mathrm{Tr}_{L-l}|\psi \rangle \langle \psi |$, where $|\psi \rangle
$ is the ground-state wave-function of Hamiltonian (\ref{H1}) and $l$ is the length of the left block for a specific bipartition. The quantum criticality of the
interaction-driven topological phase transition can be characterized with the von
Neumann entropy \cite%
{Flammia2009,Hastings2010,Daley2012,Abanin2012,jiang2012,Islam2015}
\begin{equation}
S_{\mathrm{vN}}=-\mathrm{Tr}_{l}[\hat{\rho}_{l}\log \hat{\rho}_{l}],
\end{equation}%
with $l=L/2$ the half part of the lattice.
It is believed that the property underlying the long-range correlations is
entanglement~\cite{osborne2002entanglement}, and, on the other hand, the correlation length becomes divergent at the critical point of continuous phase transition~\cite{SachdevQPT}. The divergence of the von Neumann entropy at the critical point thus indicates a continuous transition~\cite{Pollmann2010}. Besides, the von Neumann entropy also reveals the central charge of the conformal field
theory underlying the critical behavior, which
generally determines the effective field theory and reflects the universality class of phase transition~\cite{pasquale}.
For a critical system under periodic boundary conditions, the von Neumann entropy of a subchain of length $l$ scales as
\begin{equation}
S_{\mathrm{vN}}(l)=\frac{c}{3}\ln \left[ \sin \frac{\pi l}{L}\right] +\text{const},
\end{equation}%
in which, the slope at large distance gives the central charge $c$ of the
conformal field theory~\cite%
{centralcharge1,centralcharge2,centralcharge3,centralcharge4}. The formula for the case under open boundary condition is obtained by replacing $3$ by $6$.

One also can use the fidelity of the wavefunction of the ground states to identify the phase transition, which can be defined as the modulus of the overlap between two states($\psi^{'},\psi$)~\cite{fidelity}
\begin{equation}
F(\psi^{'},\psi) = |\langle \psi^{'} | \psi \rangle|,
\end{equation}%
where $|\psi\rangle$ and $|\psi^{'}\rangle$ are the input and output states respectively, and both of them are normalized.
The topological ground state of extended SSH model under periodic boundary condition
is nondegenerate and separated from the first excited state by a finite gap,
which closes and reopens across a topological phase transition. The
excited energy gap is defined as%
\begin{equation}
\Delta _{e}=E_{e}^{p}(N)-E_{g}^{p}(N),
\end{equation}%
where $E_{e}^{p}(N)$ [$E_{g}^{p}(N)$] is the first-excited (ground) state
energy of $N$ atoms under periodic boundary condition.
As we all known, the nearest-neighbor interaction $V$ can induce the CDW phase, in which the CDW order parameters can be defined as
\begin{equation}\label{CDW_order}
C= \frac{1}{L} \sum_{i=1}^{L} (-1)^i \langle \hat{n}_i \rangle.
\end{equation}
For TI under open boundary condition, the presence of localized density of edge mode is its typical feature.
The density distribution of the edge modes can be calculated as
\begin{equation}
\langle \Delta \hat{n}_j \rangle = \langle \hat{n}_j(N+1) \rangle - \langle%
\hat{n}_j(N)\rangle,
\end{equation}%
with $\langle\hat{n}_j(N)\rangle$ the density distribution for $N$ atoms
under the open boundary condition.

\begin{figure}[tb]
\centering
\includegraphics[width = 3.5in]{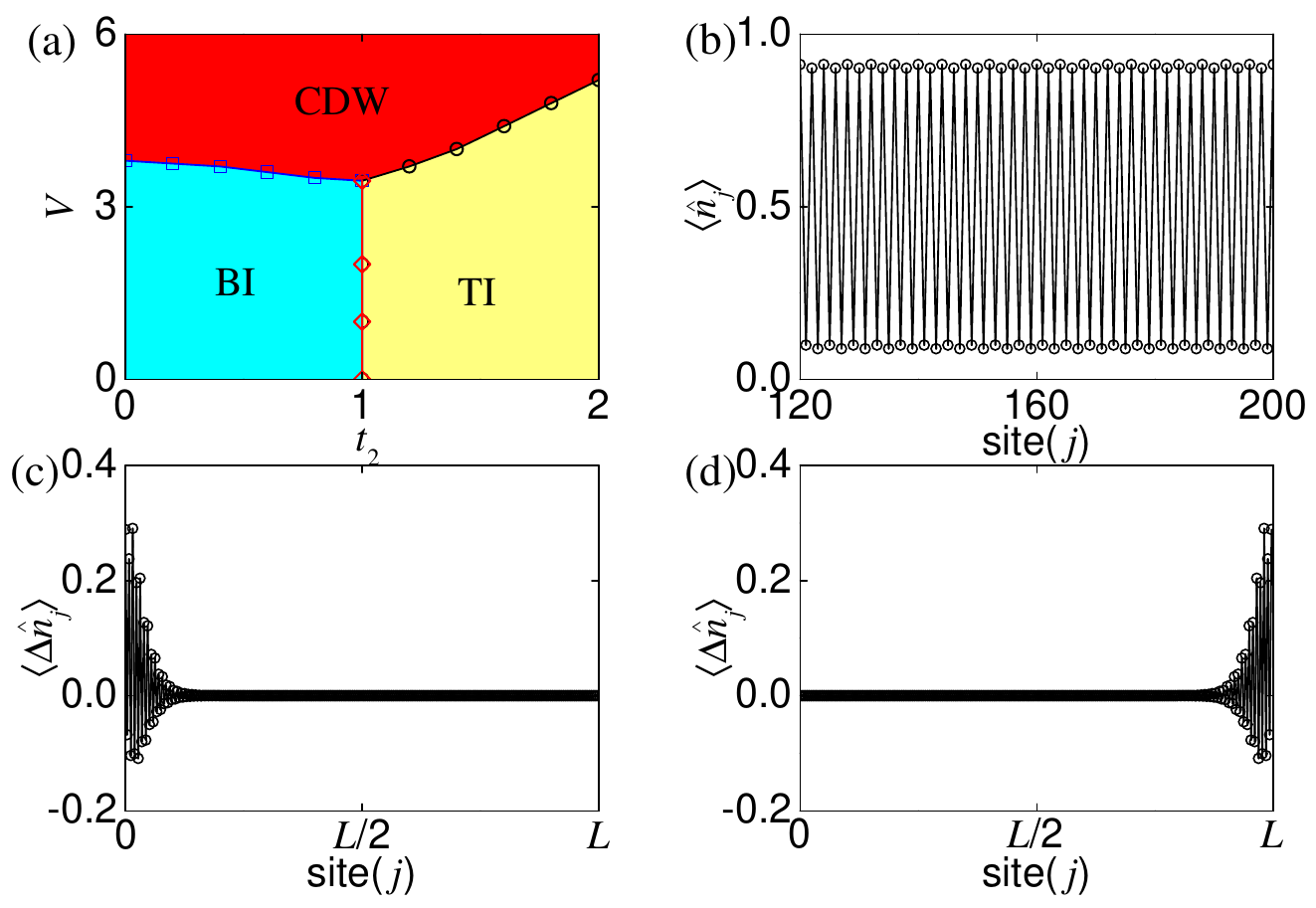} \hskip 0.0cm
\caption{(a) The phase diagram of spinless SSH4 Hamiltonian (\ref{H1}) in $t_2-V$ plane in the thermodynamic limit under periodic condition, which contains TI (topological insulator), BI (band insulator), and CDW (charge density wave). The critical line between TI and CDW (black line with circle symbol) is the Luttinger liquid with central charge $c=0.46$. The critical line between TI and BI (red line with diamond symbol) is the Luttinger liquid with central charge $c=0.46$. (b) The density profile $\langle \hat{n}_j \rangle$ of CDW with $t_2=1.2$ and $V=5.0$ under periodic boundary condition.
(c) and (d) The edge-model density distributions $\langle \Delta \hat{n}_j \rangle$ of two-fold degenerate TI with $t_2=1.6$ and $V=2.0$ under open boundary condition. In (b)-(d), we have $L=320$ and $N=160$.}
\label{fig:phase diagram spinless}
\end{figure}

\begin{figure*}[tb]
\centering
\includegraphics[width = 7.0in]{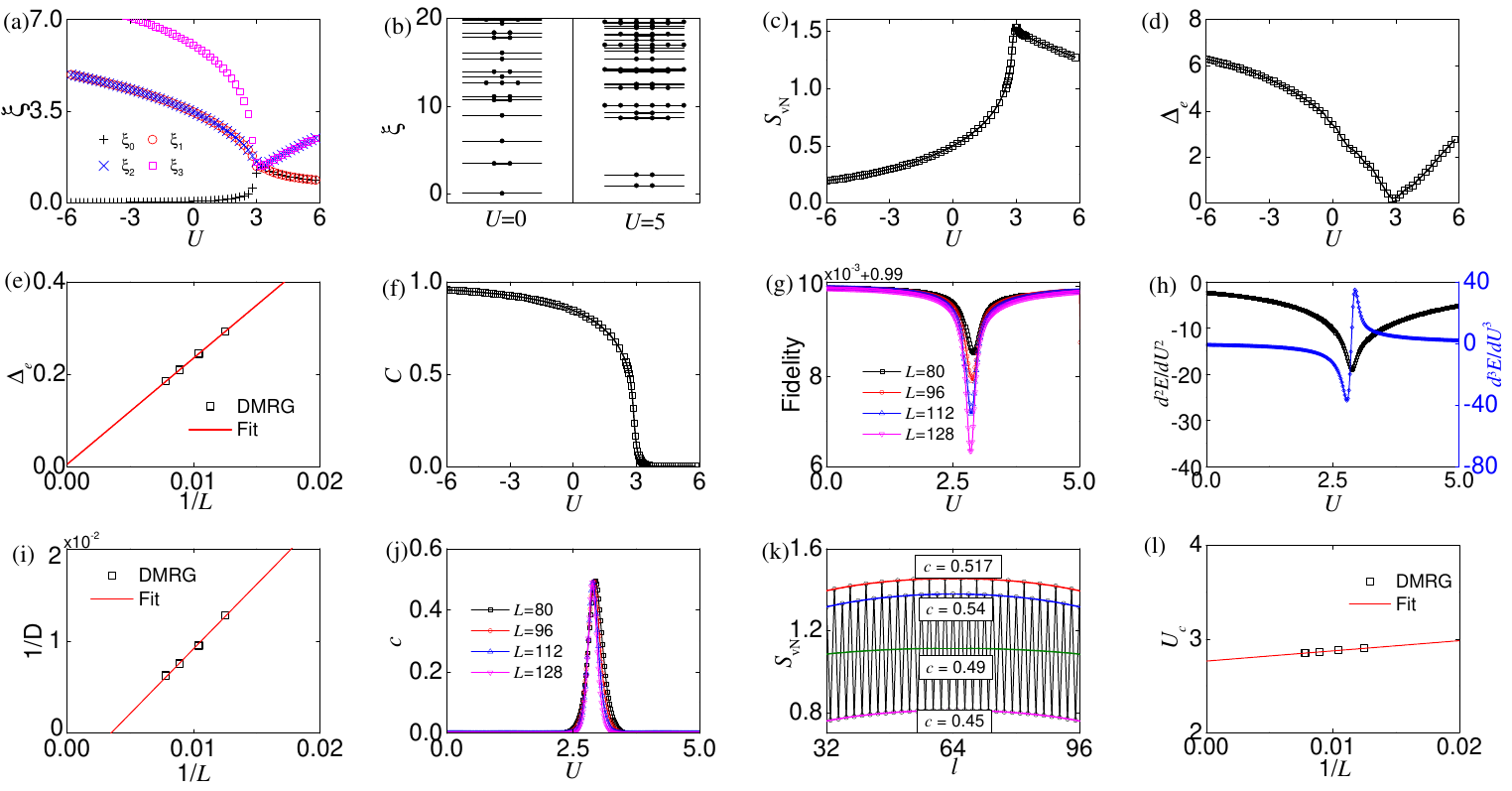} \hskip 0.0cm
\caption{Similar as the Fig.~\ref{fig:phase diagram spinless}.
(a) The entanglement spectrum $\protect\xi_{i}$ versus $U$.
(b) The several lower levels in the entanglement spectrum for two values of $U=0,5$.
(c) The von Neumann entropy $S_{\mathrm{vN}}$, and
(d) the excited energy gap $\Delta_{e}$ versus $U$.
(e) The finite-size scaling of the $\Delta_{e}$ at the critical point.
(f) The CDW order parameter $C$,
(g) the fidelity Fidelity $\langle \psi(U) | \psi(U+\delta U) \rangle$ ($\delta U=0.02$), and
(h) the derivatives of ground-state energy $d^{n}E/dU^{n}$ ($n=2,3$) versus $U$.
(i) The finite-size scaling of the inverse of peak value of the fourth-order derivative of ground-state energy, and
(j) the fitting central charge $c$ as functions of the interaction strength $U$.
(k) The scaling of the von Neumann entropy $S_{\mathrm{vN}}(l)$ as a function of subchain $l$ at the critical point.
(l) The finite-size scaling of the critical point for phase transition between TI and CDW.
In all subfigure, we have $t_2=1.6$, $V=2.0$ and $L=N=128$ except the finite-size scalings. All subfigure are under periodic boundary condition.}
\label{fig:phase_transition_2com}
\end{figure*}

\section{Many-body quantum phases}
\label{Many-body quantum phases}

\subsection{Spinless SSH4}

We first characterize the many-body properties of spinless SSH4 Hamiltonian (\ref{H1}).
Based on the experimental setup, we here fixed $t_1=t_3=t_4=1$ and vary $t_2$.
When increasing $t_2$ from $0$ to 2 in the absence of interaction, the phase is trivial band insulator (BI) when $t_2<1$,
and TI $t_2>1$ with critical point $t_2^c=1$.
Here we consider the topological phase transition driven by the nearest-neighbor
interaction $V$ for a fixed tunneling, i.e., $t_2=1.6$.
For zero and weak nearest-neighbor interaction strength $V$,
the lowest entanglement spectrum $\xi_{i}$ is two-fold
degenerate for finite lattice size $L=320$, as shown in Fig.~\ref{fig:phase transition}(a).
But the $\xi_{i}$ show the different features for noninteracting and weak interaction topological insulators.
For noninteracting topological insulator, some of the high levels of $\xi_{i}$ show the four-fold degeneracy.
However, all levels of $\xi_{i}$ for weak interacting topological insulator are two-fold degenerate
[see Fig.~\ref{fig:phase transition}(b)].
Further increasing the interaction strength $V$, $\xi_{i}$ is no longer degenerate beyond a critical interaction strength $V_{c}\sim 4.34$, as shown in Fig.~\ref{fig:phase transition}(a).
The von Neumann entropy $S_{\mathrm{vN}}$ also exhibits a sharp peak at around the critical
point, as shown in Fig.~\ref{fig:phase transition}(c).
Moreover, the excited
energy gap $\Delta_{e}$ under the periodic boundary condition closes at the critical point and then reopens,
as shown in Figs.~\ref{fig:phase transition}(d) and \ref{fig:phase transition}(e).
In this processing of the phase transition, the CDW order parameter $C$
become finite values from zero when $V$ beyond the critical strength $V_{c}$, as shown in Fig.~%
\ref{fig:phase transition}(f).
Above all, one can conclude that the nearest-neighbor interaction $V$ drives the TI into the CDW phase through a phase transition.

In the CDW phase, the chiral symmetry protecting the nontrivial topological phase has been spontaneously broken by the CDW order~\cite{Yahyavi2018}. The ground states of interacting SSH model are approximately equivalent to that of a non-interacting SSH model plus an additional on-site staggered term in the CDW phase~\cite{Yahyavi2018}. Hence, the phase transitions from the topologically trivial/nontrivial phases to the CDW phase can be classified with Landau's paradigm. Specifically, the local order parameter characterizing the phase transition is the CDW order defined in Eq. (\ref{CDW_order}). Although, the CDW phase is a topologically trivial phase, evidenced by the lacking of entanglement-entropy degeneracy, as shown in Fig.~\ref{fig:phase transition}(a). The phase transition to the CDW phase is not a standard topological phase transition in the common sense, because it can be described with local order parameters and accompanies the spontaneous breaking of symmetries. It is consistent with the transition from the topological band insulator to the antiferromagnetic Mott insulator in two-dimensional Kane-Mele-Hubbard model~\cite{hohenadler2012}, which is in the universality class of three-dimensional XY model that also accompanies spontaneous symmetry breaking~\cite{aloysius1993}.

As expected from the previous literature~\cite{fidelity}, a sharp dip on the curve of fidelity $|\langle \psi(V) | \psi(V+\delta V) \rangle|$ ($\delta V=0.02$) accompanies with the
quantum phase transition to emerge, and the dip becomes sharper and sharper as the system size increases, as shown in Fig.~\ref{fig:phase transition}(g). The behavior is associated with the dramatic change of ground state in the critical regime.
It is argued that the phase transition to CDW phase in interacting SSH model is a first-order phase transition when the difference between the alternating hopping coefficients is not too large, based on the variational study~\cite{Yahyavi2018}. In contrast, we find this transition in the interacting SSH4 model considered here is a continuous (third-order) phase transition, as directly evidenced by the discontinuousness of the third-order derivative of ground-state energy [see Fig.~\ref{fig:phase transition}(h)]. Actually, the sharp peak of entanglement entropy shown in Fig.~\ref{fig:phase transition}(c) also indicates the phase transition is continuous. Note that the discontinuousness of the third-order derivative of ground-state energy looks not so obvious in the figure is due to that the system size is finite in the numerical calculation, but it obviously shows a sharp jump at around the critical point. As shown in Fig.~\ref{fig:phase transition}(i), the finite-size scaling indicates that the fourth-order derivative of ground-state energy moves toward the divergent regime in the thermodynamic limit (the relatively large deviation is due to the derivatives are calculated with numerical difference order by order). This observation further confirms the third-order derivative of energy is not continuous and the CDW phase transition is a third-order phase transition.

At the critical point between the TI and CDW, the energy spectrum is gapless in the thermodynamic limit
[i.e. $\Delta_e = 0$; {see Fig.~\ref{fig:phase transition}(e)}] and the scaling of von Neumann entropy $S_{\mathrm{vN}}(l)=\frac{c%
}{3}\ln[\sin(\protect\pi l/L)]+\text{const}$ with a central charge $c=0.46$, as shown in Figs.~\ref{fig:phase transition}(j) and \ref{fig:phase transition}(k).
The critical line is the Luttinger liquid with central charge $c=0.46$.
By using finite-size scaling, we get the critical points of interaction-driven Landau's phase transitions between TI and CDW $V_c=4.53$ when $t_2=1.6$ in the thermodynamic limit for interacting SSH4 model, as shown in Fig.~\ref{fig:phase transition}(l).
We use similar methods to identify the critical points $V_c$ for several $t_2$.

According to the calculated degeneracy of entanglement spectrum, entanglement
entropy, energy gaps, CDW parameter order, fidelity, and derivatives of ground-state energy, we can draw the phase
diagram in the $t_2-V$ plane, as shown in Fig.~\ref{fig:phase diagram spinless}(a).
This phase diagram contains three phases such as TI, BI and CDW.
By scaling the von Neumann entropy $S_{\mathrm{vN}}(l)$ of the critical lines,
we find that the critical line between TI and CDW (BI) is the Luttinger liquid
with central charge $c=0.46$.
For large nearest-neighbor interaction strength $V$, the density profile $\langle \hat{n}_j \rangle$ of the ground-state always modulates along real lattice space with periodic 2, the corresponding phase is CDW, as shown in Fig.~\ref{fig:phase diagram spinless}(b).
For weak $V$, the ground-state is TI (BI) when $t_2>1$ ($t_2<1$).
The TI not only exhibits two-fold degenerate entanglement spectrum but also
has two-fold degenerate ground-state under the open boundary condition,
in which only one edge model occupied on one edge side for each degenerate ground-state,
as shown in Figs.~\ref{fig:phase diagram spinless}(c) and \ref{fig:phase diagram spinless}(d).
For BI, the density profile is uniform (i.e. $\langle \hat{n}_j \rangle=0.5$), with the entanglement spectrum almost completely non-degenerate, which are not shown.

\subsection{Spin-1/2 SSH4}

\begin{figure}[tb]
\centering
\includegraphics[width = 3.5in]{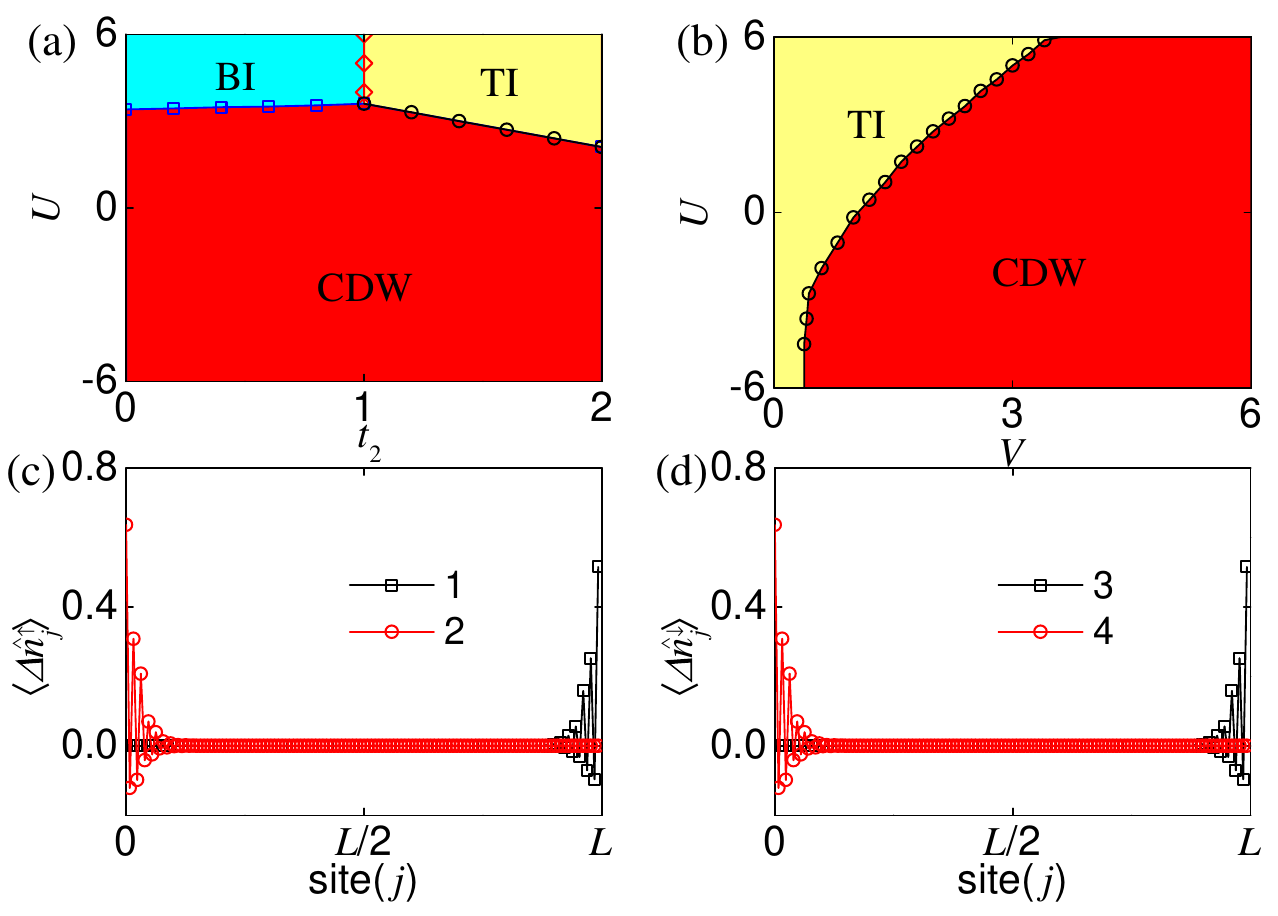} \hskip 0.0cm
\caption{The phase diagram of spin-1/2 SSH4 Hamiltonian (\ref{H2}) (a) in $t_2-U$ plane with $V=2$, and (b) in $V-U$ plane with $t_2=1.6$ in the thermodynamic limit under periodic condition. The edge-model density distributions of different spin $\langle \Delta \hat{n}_j^{\sigma} \rangle$ (c) $\sigma=\uparrow$ and (d) $\sigma=\downarrow$, with $t_2=1.6$, $U=5.0$, $V=2.0$, and $L=N=128$ under open boundary condition. In (c) and (d), the number in the figure legends label different degenerate ground-state.}
\label{fig:phase diagram}
\end{figure}

Here, we consider the spin-1/2 SSH4 Hamiltonian (\ref{H2}), which contains the on-site interaction. Similar as the Fig.~\ref{fig:phase diagram spinless}, we calculate the entanglement spectrum,
entanglement entropy, energy gaps, CDW orders, fidelity, derivatives of ground-state energy, and central charge, as shown in Fig.~\ref{fig:phase_transition_2com}.
Combing with the finite-size scaling, we map out the phase diagram of the spin-1/2 SSH4 model, as shown in Figs.~\ref{fig:phase diagram}(a) and \ref{fig:phase diagram}(b).
For the filling $N/L=1$, the atoms fully occupy the lower half of the eight energy bands in the BI and TI regime.
The repulsive on-site interaction favors less density on the same site, while the CDW phase has twice density than the uniform BI and TI cases. Therefore, repulsive on-site interaction enhances the TI (BI) phases and suppresses the CDW phase. In contrast, the attractive on-site interaction  suppresses the TI (BI) phases but enhances the CDW phase.

We would like to note that, in contrast to the spinless case, the inter-spin nearest-neighbor interaction terms in the spinful case also enhance the CDW phase, even when the on-site interaction is absent. The spinful case thus has a lower critical value of nearest-neighbor interaction for the CDW phase transition. For example, as shown in Fig.~\ref{fig:phase diagram} (Fig.~\ref{fig:phase diagram spinless}), the spinful (spinless) system already enters (have not entered) the CDW phase when $U=0$ and $V=2$ ($V=2$).

For TI in repulsive on-site interaction regime, only the one component atoms can be localized at one side of the edge.
The TI features four-fold degeneracies ground-state with the four-fold localized density distributions at the edge, such as
$\vert \! \uparrow \rangle$ located at left edge, $\vert \! \uparrow \rangle$ located at right edge,
$\vert \! \downarrow \rangle$ located at left edge, and $\vert \! \downarrow \rangle$ located at right edge,
as shown in Figs.~\ref{fig:phase diagram}(c) and \ref{fig:phase diagram}(d).
For TI in attractive on-site interaction regime, both the two component atoms with equal number localized at one side of the edge.
The ground-state are two-fold degeneracies, one is $\vert \! \uparrow + \downarrow \rangle$ located at left edge, another is $\vert \! \uparrow + \downarrow \rangle$ located at right edge.

\section{Conclusions}
\label{Conclusions}

In conclusion, we have studied theoretically the interacting
SSH4 models at half filling using DMRG method.
We find that the nearest-neighbor interaction can drive the TI (BI) to CDW phase.
Varying the tunnelings, there exist the topological phase transition between TI and BI.
The critical lines of the topological phase transitions correspond to the Luttinger liquids with integer central charges.
We have calculated
entanglement spectrum, entanglement entropy, energy gaps, and CDW order parameter, to identify the interaction-driven CDW phase transitions and phase diagrams. The phase transition to the CDW phase driven by interaction is shown to be a continuous phase transition. The central charges at the phase boundaries are fixed.
We also have studied the topological properties of the spin-1/2 interacting SSH4 model, and find that the repulsive on-site interaction can enhance the TI and suppress the CDW phase.
But the attractive on-site interaction plays the oppositive role.
We also investigate the edge-model of the TIs. In experiment, the entanglement entropy can be measured using quantum interference
of many-body twins of ultracold atoms in optical lattices~\cite%
{Islam2015}.The CDW phase can be detected by time-of-flight in cold atom experiment.
Our work provides new insights into the
many-body physics in systems with topological properties, and may stimulate the quantum simulation of strong-correlated topological insulators with cold
atoms in optical superlattices.

\section*{Acknowledgments}
X.Z. and S.J. are supported by
National Key R\&D Program of China under Grant No.
2017YFA0304203, the National Natural Science Foundation of China (NSFC) under Grant No.~12004230, the Research Project Supported by Shanxi Scholarship Council of China and Shanxi '1331KSC'. J.S.P. is supported by the NSFC under Grant No.~11904228 and the Science Specialty Program of Sichuan University under Grand No. 2020SCUNL210.
Our simulations make use of the ALPSCore library~\cite{ALPS}, based on the original ALPS project~\cite{ALPS2}.


\begin{thebibliography}{99}
\bibitem{Hasan2010} M. Z. Hasan and C. L. Kane, Colloquium: Topological insulators, Rev. Mod. Phys. \textbf{82}, 3045 (2010).
\bibitem{Qi2011} X.-L. Qi and S.-C. Zhang, Topological insulators and superconductors, Rev. Mod. Phys. \textbf{8}3, 1057 (2011).
\bibitem{Bansil2016} A. Bansil, H. Lin, and T. Das, Colloquium: Topological band theory, Rev. Mod. Phys. \textbf{88}, 021004 (2016).
\bibitem{Chiu2016} C.-K. Chiu, J. C. Y. Teo, A. P. Schnyder, and S. Ryu, Classification of topological quantum matter with symmetries, Rev. Mod. Phys. \textbf{88}, 035005 (2016).
\bibitem{Armitage2018} N. P. Armitage, E. J. Mele, and A. Vishwanath, Weyl and Dirac semimetals in three-dimensional solids, Rev. Mod. Phys. \textbf{9}0, 015001 (2018).
\bibitem{Goldman2016} N. Goldman, J. C. Budich, and P. Zoller, Topological quantum matter with ultracold gases in optical lattices, Nat. Phys. \textbf{12}, 639 (2016).
\bibitem{zhang2018} D.-W. Zhang, Y.-Q. Zhu, Y. X. Zhao, H. Yan, and S.-L. Zhu, Topological quantum matter with cold atoms, Adv. Phys. \textbf{67}, 253 (2018).
\bibitem{Cooper2019} N. R. Cooper, J. Dalibard, and I. B. Spielman, Topological bands for ultracold atoms, Rev. Mod. Phys. \textbf{91}, 015005 (2019).
\bibitem{SSH1979} W. P. Su, J. R. Schrieffer, and A. J. Heeger, Solitons in polyacetylene, Phys. Rev. Lett. \textbf{42}, 1698 (1979).
\bibitem{Atala2013} M. Atala, M. Aidelsburger, J. T. Barreiro, D. Abanin, T. Kitagawa,
E.e Demler, and I. Bloch, Direct measurement of the Zak phase in topological Bloch bands,
Nat. Phys. \textbf{9}, 795 (2013).
\bibitem{Wang2013} L. Wang, M. Troyer, and X. Dai, Topological charge pumping in a one-dimensional
optical lattice. Phys. Rev. Lett. \textbf{111}, 026802 (2013).
\bibitem{Lohse2016} M. Lohse, C. Schweizer, O. Zilberberg, M. Aidelsburger, and I. Bloch, A Thouless
quantum pump with ultracold bosonic atoms in an optical superlattice, Nat. Phys. \textbf{12}, 350 (2016).
\bibitem{Nakajima2016} S. Nakajima, T.i Tomita, S. Taie, T. Ichinose, H. Ozawa, L. Wang,
M. Troyer, and Y. Takahashi, Topological Thouless pumping of ultracold fermions, Nat. Phys. \textbf{12}, 296 (2016).
\bibitem{Leder2016} M. Leder, C. Grossert, L. Sitta, M. Genske, A. Rosch, and M. Weitz, Real-space imaging of a topologically protected edge state with ultracold atoms in an amplitude-chirped optical lattice, Nat. Comm. \textbf{7}, 13112 (2016).
\bibitem{shen2012} S. Q. Shen, Topological Insulators-Dirac Equation in Condensed Matters (Springer, New York, 2012).
\bibitem{Yoshida2014} T. Yoshida, R. Peters, S.i Fujimoto, and N. Kawakami, Characterization of a Topological Mott Insulator in One Dimension, Phys. Rev. Lett. \textbf{112}, 196404 (2014).
\bibitem{Sirker2014} J. Sirker, M. Maiti, N. P. Konstantinidis, and N. Sedlmayr, Boundary fidelity and entanglement in the symmetry protected topological phase of the SSH model, J. Stat. Mech. P10032 (2014).
\bibitem{Wang2015} D. Wang, S. Xu, Y. Wang, and C. Wu, Detecting edge degeneracy in interacting topological insulators through entanglement entropy, Phys. Rev. B \textbf{91}, 115118 (2015).
\bibitem{Liberto2016} M. Di Liberto, A. Recati, I. Carusotto, and C. Menotti, Two-body physics in the Su-Schrieffer-Heeger model, Phys. Rev. A \textbf{94}, 062704 (2016).
\bibitem{Ye2016} B.-T. Ye, L.-Z. Mu, and H. Fan, Entanglement spectrum of Su-Schrieffer-Heeger-Hubbard model, Phys. Rev. B \textbf{94}, 165167 (2016).
\bibitem{Meichanetzidis2016} K. Meichanetzidis, J. Eisert, M. Cirio, V. Lahtinen, and J. K. Pachos, Diagnosing topological edge states via Entanglement monogamy, Phys. Rev. Lett. \textbf{116}, 130501 (2016).
\bibitem{Marques2017} A. M. Marques and R. G. Dias, Multihole edge states in Su-Schrieffer-Heeger chains with interactions, Phys. Rev. B \textbf{95}, 115443 (2017)
\bibitem{Kuno2019} Y. Kuno, Phase structure of the interacting Su-Schrieffer-Heeger model and the relationship with the Gross-Neveu model on lattice, Phys. Rev. B \textbf{99}, 064105 (2019).
\bibitem{marques2018topological}A. M. Marques and R. G. Dias, Topological bound states in interacting Su--Schrieffer--Heeger rings, Journ. of Phys.: Cond. Matt. \textbf{30}, 305601 (2018).
\bibitem{GomezLeon2013} A. G\'{o}mez-Le\'{o}n and G. Platero, Floquet-bloch theory and topology in periodically driven lattices, Phys. Rev. Lett. \textbf{110}, 200403 (2013).
\bibitem{DalLago2015} V. Dal Lago, M. Atala, and L. E. F. F. Torres, Floquet topological transitions in a driven one-dimensional topological insulator, Phys. Rev. A \textbf{9}2, 023624 (2015).
\bibitem{Li2014} L. Li, Z. Xu, and S. Chen, Topological phases of generalized Su-Schrieffer-Heeger models, Phys. Rev. B \textbf{89}, 085111 (2014).
\bibitem{An2018} F. A. An, E. J. Meier, and B. Gadway, Engineering a flux-dependent mobility edge in
disordered zigzag chains, Phys. Rev. X \textbf{8}, 031045 (2018).
\bibitem{PerezGonzalez2019} B. Perez-Gonzalez, M. Bello, A. Gomez-Leon, and G. Platero, Interplay between
long-range hopping and disorder in topological systems, Phys. Rev. B \textbf{99}, 035146
(2019).
\bibitem{Ahmadi2020} N. Ahmadi, J. Abouie, and D. Baeriswyl, Topological and nontopological features of generalized Su-Schrieffer-Heeger models, Phys. Rev. B \textbf{10}1, 195117 (2020).
\bibitem{kumar2021} R. R. Kumar, N. Roy, Y. R. Kartik, S. Rahul, and S. Sarkar, Topological phase transition at quantum criticality, arXiv preprint arXiv:2112.02485 (2021).
\bibitem{zhangtwolegSSH} S.-L. Zhang and Q. Zhou, Two-leg Su-Schrieffer-Heeger chain with glide reflection symmetry, Phys. Rev. A \textbf{95}, 061601(R) (2017).
\bibitem{Sun2017} N. Sun and L.-K. Lim, Quantum charge pumps with topological phases in a Creutz ladder, Phys. Rev. B \textbf{96}, 035139 (2017).
\bibitem{Junemann2017} J. J\"{u}nemann, A. Piga, S.-J. Ran, M. Lewenstein, M. Rizzi, and A. Bermudez, Exploring interacting topological insulators with ultracold atoms: the synthetic creutz-hubbard model, Phys. Rev. X \textbf{7}, 031057 (2017).
\bibitem{Kang2018} J. H. Kang, J. H. Han, and Y. Shin, Realization of a Cross-Linked Chiral Ladder with Neutral Fermions in a 1D Optical Lattice by Orbital-Momentum Coupling, Phys. Rev. Lett. \textbf{121}, 150403  (2018).
\bibitem{He2021} J Y. He, R. Mao, H. Cai, J.-X. Zhang, Y. Li, L. Yuan, S.-Y. Zhu, and D.-W. Wang, Flat-Band Localization in Creutz Superradiance Lattices, Realization of a Cross-Linked Chiral Ladder with Neutral Fermions in a 1D Optical Lattice by Orbital-Momentum Coupling, Phys. Rev. Lett. \textbf{126},103601 (2021).
\bibitem{SSH42018} M. Maffei, A. Dauphin, F. Cardano, M. Lewenstein, and P.
Massignan, Topological characterization of chiral models through their long
time dynamics. New J. Phys. \textbf{20}, 013023 (2018).
\bibitem{Celi2014} A. Celi, P. Massignan, J. Ruseckas, N. Goldman, I. B. Spielman, G. Juzeli\={u}nas, and M. Lewenstein, Synthetic gauge fields in synthetic dimensions. Phys. Rev. Lett. \textbf{112}, 043001 (2014).
\bibitem{Price2015} H. M. Price, O. Zilberberg, T. Ozawa, I. Carusotto, and
N. Goldman, Four-dimensional quantum Hall effect with ultracold atoms, Phys.
Rev. Lett. \textbf{115}, 195303 (2015).
\bibitem{Mancini2015} M. Mancini, G. Pagano, G. Cappellini, L. Livi, M. Rider, J. Catani, C. Sias, P. Zoller, M. Inguscio, M. Dalmonte, and L. Fallani, Observation of chiral edge states with neutral fermions in synthetic Hall ribbons, Science \textbf{349}, 1510 (2015).
\bibitem{Stuhl2015} B. K. Stuhl, H.-I. Lu, L. M. Aycock, D. Genkina, and I. B. Spielman, Visualizing edge states with an atomic Bose gas in the quantum Hall regime, Science \textbf{349}, 1514 (2015).
\bibitem{Lustig2019} E. Lustig, et al. Photonic topological insulator in synthetic dimensions. Nature \textbf{56}7, 356 (2019).
\bibitem{marques2020analytical}A. M. Marques and R. G. Dias, Analytical solution of open crystalline linear 1D tight-binding models, Journ. of Phys. A: Math. and Theor. \textbf{53}, 075303 (2020).
\bibitem{SSH42019} D. Xie, W. Gou, T. Xiao, B. Gadway, and B. Yan,
Topological characterizations of an extended Su-Schrieffer-Heeger model,
npj Quantum Inf \textbf{5}, 55 (2019).
\bibitem{SSH42020} Y. He and C.-C. Chien, Non-Hermitian generalizations of
extended Su-Schrieffer-Heeger models, J. Phys.: Condens. Matter \textbf{33}, 085501 (2021).
\bibitem{dmrg1} S. R. White, Density matrix formulation for quantum
renormalization groups, Phys. Rev. Lett. \textbf{69}, 2863 (1992).
\bibitem{dmrg2} U. Schollw\"{o}k, The density-matrix renormalization group,
Rev. Mod. Phys. \textbf{77}, 259 (2005).
\bibitem{Altland1997} A. Altland and M. R. Zirnbauer, Nonstandard symmetry
classes in mesoscopic normal-superconducting hybrid structures, Phys. Rev. B
\textbf{55}, 1142 (1997).
\bibitem{Yahyavi2018} M. Yahyavi, L. Saleem and B. Het\'{e}nyi, Variational study of the interacting, spinless Su-Schrieffer-Heeger model, J. Phys.: Condens. Matter \textbf{30}, 445602 (2018).
\bibitem{Schnyder2008} A. P. Schnyder, S. Ryu, A. Furusaki, and A. W. W.
Ludwig, Classification of topological insulators and superconductors in
three spatial dimensions, Phys. Rev. B \textbf{78}, 195125 (2008).
\bibitem{Ludwig2016} A. W. W. Ludwig, Topological phases: classification of
topological insulators and superconductors of non-interacting fermions, and
beyond, Phys. Scr., T \textbf{168}, 014001 (2016).
\bibitem{Morimoto2015} T. Morimoto, A. Furusaki, and C. Mudry, Breakdown of
the topological classification $Z$ for gapped phases of noninteracting
fermions by quartic interactions, Phys. Rev. B \textbf{92}, 125104 (2015).
\bibitem{DMRG_details} Although the truncation errors (the sum of all discarded eigenvalues in the reduced density matrix) for the excited states are far larger than those for the ground states, the maximam truncation errors are basically smaller than $10^{-6}$ and are acceptible. In spinless model with lattice length $L=320$, the ground states have truncation errors smaller than $10^{-31}$. The excited states even still have truncation errors smaller than $10^{-11}$. For the spinful model with lattice length $L=128$, the truncation errors for the ground state are smaller than $10^{-9}$. Even for the excited states in the spinful model, we still have truncation errors smaller than $10^{-6}$.
\bibitem{Li2008} H. Li and F. D. M. Haldane, Entanglement spectrum as a
generalization of entanglement entropy: Identification of topological order
in non-Abelian fractional quantum Hall effect states, Phys. Rev. Lett.
\textbf{101}, 010504 (2008).
\bibitem{Zhao2015} J.-Z. Zhao, S.-J. Hu, and P. Zhang, Symmetry-protected
topological phase in a one-dimensional correlated bosonic model with a
synthetic spin-orbit coupling, Phys. Rev. Lett. \textbf{115}, 195302 (2015).
\bibitem{Turner2011} A. M. Turner, F. Pollmann, and E. Berg, Topological
phases of one-dimensional fermions: An entanglement point of view, Phys.
Rev. B \textbf{83}, 075102 (2011).
\bibitem{Pollmann2010} F. Pollmann, A. M. Turner, E. Berg, and M. Oshikawa,
Entanglement spectrum of a topological phase in one dimension, Phys. Rev. B
\textbf{81}, 064439 (2010).
\bibitem{Fidkowski2010} L. Fidkowski, Entanglement spectrum of topological
insulators and superconductors, Phys. Rev. Lett. \textbf{104}, 130502 (2010).
\bibitem{Flammia2009} S. T. Flammia, A. Hamma, T. L. Hughes, and X.-G. Wen,
Topological entanglement Renyi entropy and reduced density matrix structure,
Phys. Rev. Lett. \textbf{103}, 261601 (2009).
\bibitem{Hastings2010} M. B. Hastings, I. Gonz\'{a}lez, A. B. Kallin, and R.
G. Melko, Measuring R\'{e}nyi entanglement entropy in quantum monte carlo
simulations, Phys. Rev. Lett. \textbf{104}, 157201 (2010).
\bibitem{Daley2012} A. J. Daley, H. Pichler, J. Schachenmayer, and P.
Zoller, Measuring entanglement growth in quench dynamics of bosons in an
optical lattice, Phys. Rev. Lett. \textbf{109}, 020505 (2012).
\bibitem{Abanin2012} D. A. Abanin and E. Demler, Measuring entanglement
entropy of a generic many-body system with a quantum switch, Phys. Rev.
Lett. \textbf{109}, 020504 (2012).
\bibitem{jiang2012} H.-C. Jiang, Z.-H. Wang, and L. Balents, Identifying
topological order by entanglement entropy, Nat. Phys. \textbf{8}, 902 (2012).
\bibitem{Islam2015} R. Islam, R. Ma, P. M. Preiss, M. E. Tai, A. Lukin, M.
Rispoli, and M, Greiner, Measuring entanglement entropy in a quantum
many-body system, Nature \textbf{528}, 77 (2015).
\bibitem{osborne2002entanglement} T. J. Osborne, and M. A. Nielsen, Entanglement, quantum phase transitions, and density matrix renormalization, Quant. Inform. Proc. \textbf{1}, 45--53 (2002).
\bibitem{SachdevQPT}S. Sachdev, Quantum Phase Transitions (Cambridge University Press, Cambridge, 1999).
\bibitem{pasquale}P. Calabrese, and J. Cardy, Entanglement Entropy and Quantum Field Theory, J. Stat. Mech. 2004, P06002(2014)
\bibitem{centralcharge1} G. Vidal, J. Latorre, E. Rico, and A. Kitaev, Entanglement in Quantum Critical Phenomena, Phys. Rev. Lett. \textbf{90}, 227902 (2003).
\bibitem{centralcharge2} P. Calabrese and J. Cardy, Entanglement entropy and quantum field theory, J. Stat. Mech. 06 (2004) P06002.
\bibitem{centralcharge3} P. Calabrese and J. Cardy, Entanglement entropy and conformal field theory, J. Phys. A \textbf{42}, 504005 (2009).
\bibitem{centralcharge4} A. E. B. Nielsen, G. Sierra, and J. I. Cirac, Violation of the area law and long-range correlations in infinite-dimensional-matrix product states, Phys. Rev. A \textbf{83}, 053807 (2011).
\bibitem{hohenadler2012} M. Hohenadler, Z. Y. Meng, T. C. Lang, S. Wessel, A. Muramatsu, and F. F. Assaad, Quantum phase transitions in the Kane-Mele-Hubbard model, Phys. Rev. B \textbf{85}, 115132(2012).
\bibitem{aloysius1993}Aloysius P.Gottlob, and MartinHasenbusch, Critical behaviour of the 3D XY-model: a Monte Carlo study, Physica A: Statistical Mechanics and its Applications \textbf{201},  593--613(1993).
\bibitem{fidelity} Shi-Jian Gu, Fidelity approach to quantum phase transitions, Int. J. Mod. Phys. B \textbf{24}, 4371(2010).
\bibitem{ALPS} A. Gaenko, A. Antipov, G. Carcassi, T. Chen, X. Chen, Q. Dong, L. Gamper, J. Gukelberger, R. Igarashi, S. Iskakov, M. Konz, J. LeBlanc, R. Levy, P. Ma, J. Paki, H. Shinaoka, S. T\"{o}do, M. Troyer, and E. Gull, Updated core libraries of the ALPS project, Computer Physics Communications 213, 235 (2017).
\bibitem{ALPS2} B. Bauer, L. D. Carr, H. G. Evertz, A. Feiguin, J. Freire, S. Fuchs, L. Gamper, J. Gukelberger, E. Gull, S. Guertler, A. Hehn, R. Igarashi et al., The ALPS project release 2.0: open source software for strongly correlated systems, Journal of Statistical Mechanics: Theory and Experiment 2011(05), P05001 (2011).

\end{thebibliography}
\end{document}